\documentclass[aps,prb,reprint,superscriptaddress]{revtex4-2}

\usepackage{amsmath,amsthm,amssymb,amsfonts,bm,graphicx,hyperref}
\hypersetup{
    colorlinks = true,
    linkcolor = red,
    urlcolor = ,
    citecolor = red,
}

\newcommand{\bra}[1]{\langle#1|}
\newcommand{\ket}[1]{|#1\rangle}
\newcommand{\bk}[2]{\left\langle#1\middle|#2\right\rangle}

\begin{document}


\title{Non-Bloch band theory of nonlinear eigenvalue problems}


\author{Kota Otsuka}
\affiliation{Institute for Solid State Physics, University of Tokyo, Kashiwa, Chiba 277-8581, Japan}
\author{Kazuki Yokomizo}
\affiliation{Department of Physics, The University of Tokyo, 7-3-1 Hongo, Bunkyo-ku, Tokyo, 113-0033, Japan}


\date{\today}

\begin{abstract}

Nonlinear eigenvalue problems arise in a wide range of physical systems, in which system parameters depend on the eigenvalue. Such systems have been proposed to exhibit an extreme sensitivity of their spectra to boundary conditions, which leads to the breakdown of conventional topological characterizations. In this work, we establish a non-Bloch framework for calculating continuum bands that reproduce the spectra of the nonlinear system with open boundary conditions. This non-Bloch band theory enables us not only to calculate the eigenvalues but also to reveal phenomena unique to the nonlinear system. We further investigate the topological bulk-boundary correspondence in a nonlinear Chern insulator within an extended version of this framework.

\end{abstract}


\maketitle

\section{Introduction}
Nonlinear dynamics gives rise to various intriguing phenomena due to the breakdown of the superposition principle~\cite{Strogatz2024}, and appears, for example, in optics~\cite{Kivshar2003,Smirnova2020} and weakly interacting bosonic systems~\cite{Gross1961,Pitaevskii1961}. These systems are described by eigenvalue equations in which the generator of time evolution depends on the eigenstate. In recent years, there has grown interest in how the nonlinearity affects topological systems and non-Hermitian systems. For instance, the bulk-boundary correspondence has been established in terms of nonlinear topological invariants~\cite{Tuloup2020,Zhou2022,Sone2024}, and novel nonlinear phenomena utilizing topological edge states have been proposed~\cite{Harari2018,Bandres2018,Sone2025,Castro2025}. In addition, lasing modes can be excited through the localization of numerous eigenstates induced by the non-Hermitian skin effect~\cite{Ezawa2022,Teo2022,Zhu2022,Leefmans2024}.

Meanwhile, nonlinear eigenvalue equations~\cite{Guttel2017,Voss2014}, in which the generator of time evolution depends on the eigenvalue, have also attracted much attention. These equations describe a variety of physical systems, such as mechanical oscillators~\cite{Fang2006,Yao2008,Huang2009,Huang2011,Lee2016,Srikanth2026}, electrical circuits~\cite{Yang2024,Zhang2026}, and dispersive metamaterials~\cite{Kittel1996,Jackson1999,Kuzmiak1994,Smith2002,Toader2004,Raghu2008,Mai2022}. Recent theoretical studies have shed light on topological aspects of this type of nonlinear systems. A pioneering work has established the bulk-boundary correspondence in terms of topology of an auxiliary system~\cite{Isobe2024}, which is analogous to that of Hermitian systems. Subsequent studies have shown that the auxiliary eigenvalue approach can be applied to investigate several other models~\cite{Cheng2024,Bai2025,Bai2025_2}. Furthermore, Refs.~\cite{Yoshida2025,Ma2026} have investigated complex-valued eigenvalues by analogy with non-Hermitian topology~\cite{Gong2018,Kawabata2019_2,Zhang2020,Okuma2020}, providing topological characterizations of the skin effect. In both cases, the Bloch band description provides a useful way to obtain the spectra of the nonlinear systems.

The intriguing feature of the nonlinear system is its extreme sensitivity to boundary conditions. Indeed, spatial nonreciprocity has been proposed to induce the discrepancy between the spectra under periodic and open boundary conditions (PBCs and OBCs)~\cite{Yoshida2025,Ma2026}. In other words, the conventional Bloch band theory fails to describe the spectra under OBCs, which complicates the analysis of the nonlinear system. For example, it is hard to calculate the OBC spectra, since solving the nonlinear eigenvalue problem is challenging even numerically. In addition, topological edge states are difficult to access due to the breakdown of the bulk-boundary correspondence~\cite{Ma2026}. Therefore, it is necessary to establish a systematic framework for calculating bands that reproduce the spectra of the nonlinear system with OBCs.

In this paper, we develop a theoretical framework for calculating the continuum bands of tight-binding systems described by the nonlinear eigenvalue equations. The key idea is that this construction is analogous to that of the non-Bloch band theory of non-Hermitian systems~\cite{Yao2018,Yokomizo2019,Yokomizo2020}. Within this framework, we study several one-dimensional (1D) models and demonstrate that the continuum bands obtained from the generalized Brillouin zones (GBZs) reproduce the spectra under OBCs. We further show that the nonlinearity induces the skin effect in a system with nonlinear pseudo-Hermiticity. Additionally, we find that the present non-Bloch framework can be extended to two-dimensional (2D) tight-binding systems with certain symmetry. We then establish the bulk-boundary correspondence between a Chern number defined from the GBZs of auxiliary bands and the emergence of topological edge states.

The rest of this paper is organized as follows. In Sec.~\ref{general_framework}, we introduce a general framework for the nonlinear eigenvalue problem and construct the corresponding non-Bloch band theory. In Sec.~\ref{1Dsystems}, we present a simple model and explain how to calculate the continuum bands. We further investigate the localization behavior of a model with nonlinear pseudo-Hermiticity.  In Sec.~\ref{2Dsystems}, we introduce 2D nonlinear systems to which the non-Bloch band theory investigated here can be applied, and establish the bulk-boundary correspondence in a Chern insulator with nonreciprocal hopping amplitudes. In Sec.~\ref{summary_outlook}, we summarize our results and discuss future perspectives.

\section{general framework\label{general_framework}}

In this section, we consider a 1D tight-binding system whose Hamiltonian exhibits the eigenvalue dependence. We then construct a theoretical framework for calculating the continuum bands that reproduce the spectra of the system with OBCs.

\subsection{Nonlinear eigenvalue problem}

We consider a 1D tight-binding system described by a nonlinear eigenvalue equation
\begin{equation}
    H(\omega)\ket{\psi(\omega)}=f(\omega)\ket{\psi(\omega)},\label{rs_nep}
\end{equation}
where $\omega\in\mathbb{C}$ is the eigenvalue, $H(\omega)$ is the eigenvalue-dependent Hamiltonian, and $f(\omega)$ is an analytical function of $\omega$. Let $L$ and $q$ denote the system size and the number of internal degrees of freedom per unit cell, respectively. The eigenvalue-dependent Hamiltonian can then be expressed as
\begin{eqnarray}
    &&H(\omega)=\bm{c}^\dagger A(\omega)\bm{c},\label{H_rs}\\
    &&\bm{c}=(c_{1,1},c_{1,2},\ldots,c_{1,q},\ldots,c_{L,1},c_{L,2},\ldots,c_{L,q})^\mathrm{T},
\end{eqnarray}
where $c_{n,\mu}$ represents the annihilation operator for the $\mu$th internal degree of freedom at the $n$th unit cell, and $A(\omega)$ is a $qL\times qL$ eigenvalue-dependent matrix. The eigenvalues of Eq.~\eqref{rs_nep} can be calculated by solving the equation given by
\begin{equation}
    \det M(\omega)=0.\label{rs_eq}
\end{equation}
Here, $M(\omega)$ is defined as
\begin{equation}
    M(\omega):=A(\omega)-f(\omega)\bm{1}_{qL},\label{M_rs}
\end{equation}
where $\bm{1}_{qL}$ is a $qL\times qL$ identity matrix. We note that, when the matrix $A(\omega)$ is independent of $\omega$ and $f(\omega)=\omega$, the above procedure reduces to solving a conventional eigenvalue equation.

When the system size is relatively large, it is hard to solve the nonlinear eigenvalue equation \eqref{rs_nep} both analytically and numerically. The conventional Bloch band theory allows us to readily obtain the spectra of the nonlinear system with PBCs~\cite{Isobe2024}. However, the nonlinear system exhibits the extreme sensitivity to boundary conditions; the discrepancy between the Bloch bands and the OBC spectra has been observed~\cite{Yoshida2025,Ma2026}. Thus, it is necessary to construct a general theory to calculate bands that reproduce the OBC spectra.

\subsection{Non-Bloch band theory for nonlinear systems\label{nB_th}}
We construct a theoretical framework for calculating the continuum bands of the nonlinear system, which is analogous to the non-Bloch band theory of non-Hermitian systems~\cite{Yao2018,Yokomizo2019,Yokomizo2020}. We rewrite the eigenvalue-dependent Hamiltonian \eqref{H_rs} as
\begin{equation}
    H(\omega)=\sum_{n}\sum_{i=-N}^N\sum_{\mu,\nu=1}^qt_{i,\mu\nu}(\omega)c_{n+i,\mu}^\dagger c_{n,\nu},\label{fff}
\end{equation}
where particles can hop up to the $N$th nearest unit cell, and $t_{i,\mu\nu}(\omega)$ are the eigenvalue-dependent hopping amplitudes toward the $i$th nearest unit cell. The eigenstate in Eq.~\eqref{rs_nep} is represented by
\begin{equation}
    \ket{\psi(\omega)}=(\psi_{1,1}(\omega),\ldots,\psi_{L,q}(\omega))^\mathrm{T}.\label{ddd}
\end{equation}
Building on a general theory of linear difference equations, for a given $\omega$, the eigenstate components can be written as a linear combination,
\begin{equation}
    \psi_{n,\mu}(\omega)=\sum_j\beta_j^{-n}\phi_\mu^{(j)}(\omega).\label{translaw}
\end{equation}
Here, the set of $(\beta_j,\phi_\mu^{(j)}(\omega))$ is determined by the eigenvalue equation
\begin{equation}
        \sum_{\nu=1}^q[\mathcal{H}(\omega,\beta)]_{\mu\nu}\phi_\nu(\omega)=f(\omega)\phi_\mu(\omega),\label{k_ch_eq}
\end{equation}
where
\begin{equation}
    [\mathcal{H}(\omega,\beta)]_{\mu\nu}:=\sum_{i=-N}^Nt_{i,\mu\nu}(\omega)\beta^{i}.\label{nB_H}
\end{equation}
The condition for the nontrivial solutions of eigenvalue equation \eqref{k_ch_eq} is then given by
\begin{equation}
    \det\mathcal{M}(\omega,\beta)=0.\label{M_ch_eq}
\end{equation}
Here, the matrix $\mathcal{M}(\omega,\beta)$ is defined as
\begin{equation}
    \mathcal{M}(\omega,\beta):=\mathcal{H}(\omega,\beta)-f(\omega)\bm{1}_{q},\label{M_k}
\end{equation}
where $\bm{1}_{q}$ is a $q\times q$ identity matrix. We note that Eq.~\eqref{M_ch_eq} is an algebraic equation for $\beta$ of $2qN$th degree.

By analogy with the conventional non-Bloch band theory, the solutions of Eq.~\eqref{M_ch_eq} form continuum sets of $\beta=e^{ik}$, where $k$ is the complex-valued Bloch wavenumber, and these sets allow for obtaining the continuum bands. The continuum sets of $k$ determine the trajectories on the $\beta$-complex plane, which are referred to as the GBZs. For the solutions of Eq.~\eqref{M_ch_eq}, the condition for the GBZs is given by
\begin{equation}
    |\beta_{qN}|=|\beta_{qN+1}|,\label{GBZ}
\end{equation}
where the solutions satisfy
\begin{equation}
    |\beta_1|\le|\beta_2|\le\cdots\le|\beta_{2qN-1}|\le|\beta_{2qN}|.\label{eee}
\end{equation}
In Appendix \ref{AppA}, we show that the OBC spectra under the condition \eqref{GBZ} form the continuum sets in the limit of a large system size. The continuum bands obtained by combining Eqs.~\eqref{M_ch_eq} and \eqref{GBZ} reproduce the OBC spectra. We note that solving Eq.~\eqref{M_ch_eq} with $\beta$ replaced by $e^{ik}~(k\in\mathbb{R})$ leads to the Bloch bands, which reproduce the PBC spectra.

\subsection{Skin effect\label{skin_effect}}

The non-Bloch band theory allows for determining the GBZs formed by $\beta=e^{ik}~(k\in{\mathbb C})$. Physically, the complex-valued Bloch wavenumber means that the bulk eigenstates are localized at the edges of the system. This localization phenomenon is referred to as the skin effect, analogous to the non-Hermitian skin effect in non-Hermitian systems. According to Ref.~\cite{Yoshida2025}, the skin effect in the nonlinear system can be characterized by a $\mathbb{Z}$ topological invariant defined from the real-valued Bloch wavenumber. This topological invariant is given by
\begin{equation}
    W(\omega_\text{ref}):=\frac{1}{2\pi i}\int_0^{2\pi}\mathrm{d}k\frac{\mathrm{d}}{\mathrm{d}k}\ln\det\mathcal{M}(\omega_\text{ref},e^{ik}),\label{windnum}
\end{equation}
where $\omega_\text{ref}\in\mathbb{C}$ is a reference point satisfying $\det\mathcal{M}(\omega_\text{ref},e^{ik})\neq0$. This quantity is called the winding number, and its value remains the same unless the reference point crosses the Bloch band. The nonzero winding number coincides with the localization of the bulk eigenstates, and, for example, for regions with $W(\omega_\text{ref})>0$, the bulk eigenstates are localized at the right boundary of the system. We describe the correspondence between nonzero values of the winding number and the emergence of the skin modes in Appendix~\ref{AppB1}.

\section{One-dimensional system\label{1Dsystems}}

In this section, we explore the band structures of 1D tight-binding systems described by the nonlinear eigenvalue equation \eqref{rs_nep}. We first focus on a simple model and demonstrate that the continuum bands reproduce the OBC spectra. We next investigate a system with nonlinear pseudo-Hermiticity and find that the nonlinearity can induce the skin effect. The non-Bloch band theory then enables us to analyze the distinctive localization behavior.

\subsection{Nonreciprocal hopping system\label{1DtoyModel}}

We consider a 1D nonlinear system with nonreciprocal hopping amplitudes, which has been introduced in Ref.~\cite{Yoshida2025}. The Hamiltonian of the system is given by
\begin{eqnarray}
    H(\omega)&=&\sum_{n=1}^L\left(t_\mathrm{L}c_{n,1}^\dagger c_{n,2}+t_\mathrm{R}c_{n,2}^\dagger c_{n,1}+\tanh\omega\: c_{n,2}^\dagger c_{n,2}\right)\nonumber\\&&+\sum_{n=1}^{L-1}\left(t_\mathrm{L}c_{n,2}^\dagger c_{n+1,1}+t_\mathrm{R}c_{n+1,1}^\dagger c_{n,2}\right),\label{1D_H}
\end{eqnarray}
where $t_\mathrm{L}$ and $t_\mathrm{R}$ represent the hopping amplitudes toward the left and right directions, respectively, and they take positive real values. The system involves the eigenvalue-dependent on-site potential only on the sublattice $\mu=2$. Suppose that the function $f(\omega)$ in Eq.~\eqref{rs_nep} is given by $f(\omega)=-\omega^2$. We note that the spectra of the open-boundary system are calculated by solving Eq.~\eqref{rs_eq}.

\begin{figure}[t]
    \centering
    \resizebox{\linewidth}{!}{\input{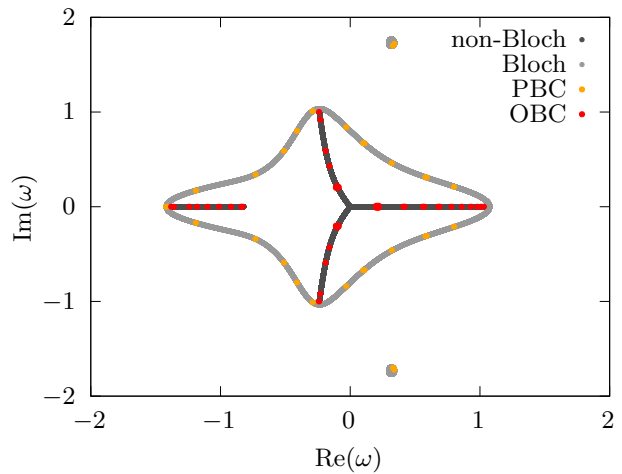}}
    \caption{Eigenvalues of the nonreciprocal hopping system. The black and gray curves represent the non-Bloch and Bloch bands, respectively. The red and orange dots indicate the spectra of the open-boundary and periodic-boundary systems with $L=10$, respectively. The parameters are set to be $t_\mathrm{R}=1$ and $t_\mathrm{L}=0.5$.}
    \label{1DtoyModel_spectra}
\end{figure}

Based on the non-Bloch framework discussed in Sec.~\ref{nB_th}, we describe the way to calculate the continuum bands from determining the GBZ. Since Eq.~\eqref{M_k} is given by
\begin{equation}
        \mathcal{M}(\omega,\beta)=\begin{pmatrix}
        \omega^2&t_\mathrm{L}+t_\mathrm{R}\beta\\
        t_\mathrm{L}\beta^{-1}+t_\mathrm{R}&\omega^2+\tanh\omega
        \end{pmatrix},
\end{equation}
the characteristic equation \eqref{M_ch_eq} can be written as a quadratic equation for $\beta$,
\begin{equation}
    t_\mathrm{R}^2\beta+2t_\mathrm{L}t_\mathrm{R}-\omega^2(\omega^2+\tanh\omega)+t_\mathrm{L}^2\beta^{-1}=0.\label{ch_eq_simple}
\end{equation}
Let $\beta_1$ and $\beta_2$ denote the solutions of Eq.~\eqref{ch_eq_simple}. The GBZ can then be constructed by $|\beta_1|=|\beta_2|$ (cf. Eq.~\eqref{GBZ}). Since $\beta_1\beta_2=(t_\mathrm{L}/t_\mathrm{R})^2$ by the Vieta's formula, the trajectory of the GBZ on the $\beta$-complex plane is determined as
\begin{equation}
    \beta=\frac{t_\mathrm{L}}{t_\mathrm{R}}e^{i\mathrm{Re}(k)}.\label{ccc}
\end{equation}
This enables us to calculate the continuum bands. Indeed, by substituting Eq.~\eqref{ccc} into Eq.~\eqref{ch_eq_simple}, we obtain the equation
\begin{equation}
    2t_\mathrm{L}t_\mathrm{R}(1+\cos\mathrm{Re}(k))-\omega^2(\omega^2+\tanh\omega)=0,\label{nB_parametric}
\end{equation}
and it determines the continuum bands (black curves in Fig.~\ref{1DtoyModel_spectra}). We also calculate the spectra of the open-boundary system (red dots in Fig.~\ref{1DtoyModel_spectra}). The key observation is that the continuum bands reproduce the OBC spectra.

To compare the non-Bloch framework with the conventional Bloch band theory, we focus on Eq.~\eqref{ch_eq_simple} with $\beta$ replaced by $e^{ik}~(k\in\mathbb{R})$. This equation is explicitly given by
\begin{equation}
    t_\mathrm{R}^2e^{-ik}+2t_\mathrm{L}t_\mathrm{R}+t_\mathrm{L}^2e^{ik}-\omega^2(\omega^2+\tanh\omega)=0.\label{B_parametric}
\end{equation}
We show the calculation result (gray curve in Fig.~\ref{1DtoyModel_spectra}) and the spectra of the periodic-boundary system (orange dots in Fig.~\ref{1DtoyModel_spectra}). These results demonstrate that the Bloch band reproduces the PBC spectra, in contrast to the non-Bloch bands. We note that the continuum bands are distributed in the region enclosed by the Bloch bands, where the winding number~\eqref{windnum} takes $+1$~\cite{Yoshida2025}.

\subsection{Nonlinearity-induced skin effect\label{NLSE}}

In this section, we investigate a localization phenomenon unique to the nonlinear system. We then consider a system that preserves the nonlinear pseudo-Hermiticity~\cite{Yoshida2025} defined as
\begin{equation}
    M(\omega^\ast)=UM(\omega)^\dagger U,\label{npH_rs}
\end{equation}
where $M(\omega)$ is given in Eq.~\eqref{M_rs}, and $U$ is a unitary and Hermitian matrix. Since the matrix $M(\omega)$ satisfies
\begin{equation}
    \det M(\omega^\ast)=\det M(\omega)^\dagger=(\det M(\omega))^\ast,
\end{equation}
the eigenvalues are restricted to either be real or appear in complex-conjugate pairs. We find that the nonlinearity induces the skin effect, and the skin modes are localized at both ends of the system. The non-Bloch band theory enables us to reveal this localization behavior by calculating the GBZs.

\subsubsection{Model\label{npH_model}}

We consider a nonlinear system whose Hamiltonian is given by
\begin{equation}
    H(\omega)=\sum_{n=1}^L{\bm c}_n^\dagger V(\omega){\bm c}_n+\sum_{n=1}^{L-1}({\bm c}_n^\dagger T{\bm c}_{n+1}+{\bm c}_{n+1}^\dagger T^\dagger{\bm c}_n),\label{Hampse}
\end{equation}
where
\begin{equation}
    V(\omega)=\begin{pmatrix}
        m+2\delta\omega&\delta\omega\\
        \delta\omega&m
    \end{pmatrix},\quad T=\begin{pmatrix}
        -t&-iv\\
        -iv&t
    \end{pmatrix},
\end{equation}
and $\bm{c}_n=(c_{n,1},c_{n,2})^\mathrm{T}$. We set $f(\omega)=\omega$ in Eq.~\eqref{rs_nep}. Suppose that all the parameters take positive real values, and $m<2\min(v,t)$. We note that the parameter $\delta$ characterizes the strength of the nonlinearity. We also remark that this system preserves the nonlinear pseudo-Hermiticity \eqref{npH_rs}, with $U$ being an identity matrix. The characteristic equation \eqref{M_ch_eq} of this system is then given by
\begin{equation}
    \det[V(\omega)+\beta^{-1}T+\beta T^\dagger-\omega\bm{1}_2]=0,\label{bbb}
\end{equation}
and it is a quartic equation for $\beta$. As a result, the condition for the GBZs is obtained by
\begin{equation}
    |\beta_2|=|\beta_3|,\label{beta2beta3}
\end{equation}
where the solutions of Eq.~\eqref{bbb} satisfy $|\beta_1|\leq|\beta_2|\leq|\beta_3|\leq|\beta_4|$.

\begin{figure}[t]
    \centering
    \resizebox{\linewidth}{!}{\input{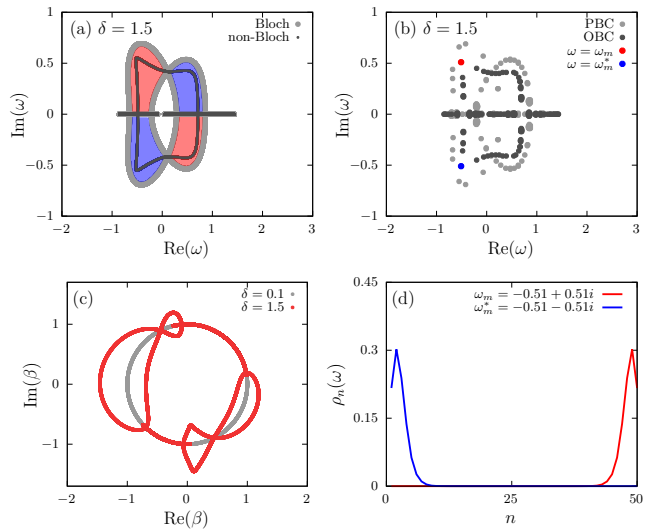}}
    \caption{Eigenvalues, generalized Brillouin zones, and eigenstates of the model with the nonlinear pseudo-Hermiticity. (a) Bloch and non-Bloch bands. The winding number \eqref{windnum} takes $+1$, $-1$, and $0$ in the red, blue, and white regions, respectively. (b) Spectra of the open-boundary and periodic-boundary systems. (c) Generalized Brillouin zones. The gray curve represents the unit circle. (d) Spatial profiles of the eigenstates defined in Eq.~\eqref{SP}. The corresponding eigenvalues are shown in the same colors in (b). The parameters are $m=0.9,t=1$, and $v=0.5$, and the system size is $L=50$ in (b) and (d).}
    \label{NLinducedNHSE}
\end{figure}

When $\delta=0$, the governing equation of the system reduces to a conventional eigenvalue equation with a Hermitian Hamiltonian. This indicates that all the eigenvalues of the system remain real for relatively small values of $\delta$, and the GBZ obtained in Eq.~\eqref{beta2beta3} forms a unit circle on the $\beta$-complex plane (gray curve in Fig.~\ref{NLinducedNHSE}(c)). As $\delta$ increases, the transition from real- to complex-valued eigenvalues occurs. By solving Eq.~\eqref{bbb} with $\beta$ replaced by $e^{ik}~(k\in\mathbb{R})$, we find that the complex-valued eigenvalues emerge when $\delta>\sqrt{2}-1$.

The complex-valued continuum bands are calculated by combining Eqs.~\eqref{bbb} and \eqref{beta2beta3} (black curves in Fig.~\ref{NLinducedNHSE}(a)). We then confirm that the continuum bands reproduce the OBC spectra (black dots in Fig.~\ref{NLinducedNHSE}(b)). For the sake of comparison, we also show the Bloch bands (gray curves in Fig.~\ref{NLinducedNHSE}(a)), which reproduces the PBC spectra (gray dots in Fig.~\ref{NLinducedNHSE}(b)). The continuum bands deviate from the Bloch bands, which means that the winding number \eqref{windnum} takes nonzero values in this system. Therefore, we find that the nonlinearity can induce the skin effect under the nonlinear pseudo-Hermiticity.

\subsubsection{Localization property}

We show the GBZs of this system in Fig.~\ref{NLinducedNHSE}(c). As $\delta$ increases, the GBZs deviate from the unit circle, which indicates the emergence of the nonlinearity-induced skin effect. We point out that one trajectory lies inside the unit circle, while the other lies outside it. Physically, this means that the bulk eigenstates are localized at both ends of the system. In Fig.~\ref{NLinducedNHSE}(d), we confirm this localization behavior by calculating the spatial profile of the bulk eigenstate,
\begin{equation}
\rho_n(\omega):=|\psi_{n,1}(\omega)|^2+|\psi_{n,2}(\omega)|^2.\label{SP}
\end{equation}
We note that the eigenstates in Fig.~\ref{NLinducedNHSE}(d) share a common localization length, as verified through the calculation of the GBZs.

Meanwhile, the bidirectional localization behavior can also be understood in terms of the winding number \eqref{windnum}. Figure~\ref{NLinducedNHSE}(a) shows that the winding number takes $+1$ and $-1$ in the red and blue regions, respectively. This indicates that the skin modes corresponding to $W=+1$ and $W=-1$ are localized at the right and left ends of the system, respectively. In Appendix~\ref{AppB2}, we show that, for a given eigenvalue $\omega$, when the eigenstate $\ket{\psi(\omega)}$ is localized at one end of the system, the localized eigenstate $\ket{\psi(\omega^\ast)}$ emerges at the opposite end.

\section{Two-dimensional system\label{2Dsystems}}

In this section, we study a 2D tight-binding system in which the skin effect occurs only in one direction due to, for example, mirror symmetry. We find that this property enables us to calculate the continuum bands within the non-Bloch framework constructed in Sec.~\ref{general_framework}. To demonstrate the validity of this approach, we consider a nonlinear Chern insulator with nonreciprocal hopping amplitudes. We also explore the bulk-boundary correspondence in terms of a Chern number defined from the GBZs of auxiliary bands.

\subsection{Non-Bloch framework\label{2d_NBF}}

We consider a 2D tight-binding system on a square lattice and assume that particles hop only in the $x$ and $y$ direction. Let $c_{(n_x,n_y),\mu}$ denote the annihilation operator for the $\mu$th internal degree of freedom at the site $(n_x,n_y)$. The Hamiltonian in the nonlinear eigenvalue equation \eqref{rs_nep} is then given by
\begin{eqnarray}
        H(\omega)&=&\sum_{n_x,n_y}\sum_{\mu,\nu=1}^q\Bigg(\sum_{i=-N_x}^{N_x}t_{i,\mu\nu}^{(x)}(\omega)c_{(n_x+i,n_y),\mu}^\dagger c_{(n_x,n_y),\nu} \nonumber\\
        &&+\sum_{i=-N_y}^{N_y}t_{i,\mu\nu}^{(y)}(\omega)c_{(n_x,n_y+i),\mu}^\dagger c_{(n_x,n_y),\nu}\Bigg),
\end{eqnarray}
where $N_x$ and $N_y$ represent the hopping range, $q$ is the number of internal degrees of freedom, and $t^{(x)}_{i,\mu\nu}(\omega)$ and $t^{(y)}_{i,\mu\nu}(\omega)$ represent the eigenvalue-dependent hopping amplitudes in the $x$ and $y$ directions, respectively. The eigenstate of Eq.~\eqref{rs_nep} is represented by
\begin{equation}
        \ket{\psi(\omega)}=(\dots,\psi_{(n_x,n_y),1}(\omega),\dots,\psi_{(n_x,n_y),q}(\omega),\dots)^{\rm T}.\label{es2D}
\end{equation}
Similar to the 1D case, for a given $\omega$, the eigenstate component in Eq.~\eqref{es2D} can be constructed by a linear combination,
\begin{equation}
    \psi_{(n_x,n_y),\mu}(\omega)=\sum_{j_x,j_y}(\beta_{x,j_x})^{-n_x}(\beta_{y,j_y})^{-n_y}\phi_{\mu}^{(j_x,j_y)}(\omega).
\end{equation}
Here, the set of $(\phi_{\mu}^{(j_x,j_y)}(\omega),\beta_x,\beta_y)$ is determined by the eigenvalue equation
\begin{eqnarray}
        &&\sum_{\nu=1}^q[\mathcal{H}_{\rm2D}(\omega,\beta_x,\beta_y)]_{\mu\nu}\phi_\nu(\omega)=f(\omega)\phi_\mu(\omega),\label{2D_k_ch_eq}\\
        &&[\mathcal{H}_{\rm2D}(\omega,\beta_x,\beta_y)]_{\mu\nu}=\sum_{i=-N_x}^{N_x}t_{i,\mu\nu}^{(x)}(\omega)\beta_x^i+\sum_{i=-N_y}^{N_y}t_{i,\mu\nu}^{(y)}(\omega)\beta_y^i.\nonumber\\
        \label{2d_ch_eq}
\end{eqnarray}
We note that solving the eigenvalue equation \eqref{2D_k_ch_eq} can be mapped to finding the solutions of the characteristic equation
\begin{equation}
    \det\mathcal{M}(\omega,\beta_x,\beta_y)=0,\label{2d_eq}
\end{equation}
where the matrix $\mathcal{M}(\omega,\beta_x,\beta_y)$ is defined as
\begin{equation}
    \mathcal{M}(\omega,\beta_x,\beta_y):=\mathcal{H}_{\rm2D}(\omega,\beta_x,\beta_y)-f(\omega)\bm{1}_q.\label{2d_M_k}
\end{equation}

By substituting $(\beta_x,\beta_y)=(e^{ik_x},e^{ik_y})~(k_x,k_y\in\mathbb{R})$ into Eq.~\eqref{2d_eq}, we obtain the Bloch bands that reproduce the spectra under full PBCs. In contrast, it is generally unclear how to calculate the continuum bands of the 2D system that reproduce the spectra under full OBCs. Nevertheless, we find that, in the system where the skin effect occurs only in one direction, the condition for the GBZs can be written in the same form as in the 1D case. A similar simplification of the condition has been studied in non-Hermitian systems~\cite{Yokomizo2023}. In the following, we consider mirror symmetry defined as
\begin{equation}
    \mathcal{U}\mathcal{M}(\omega,e^{ik_x},e^{-ik_y})^\mathrm{T}\mathcal{U}=\mathcal{M}(\omega,e^{ik_x},e^{ik_y}),\label{y_sym}
\end{equation}
where $\mathcal{U}$ is a unitary and Hermitian matrix. Physically, this symmetry suppresses nonreciprocal hopping amplitudes in the $y$ direction, which indicates that the skin effect can occur only in the $x$ direction. Therefore, one can calculate the continuum bands by combining the GBZs and Eq.~\eqref{2d_eq} with $\beta_y$ replaced by $e^{ik_y}~(k_y\in\mathbb{R})$. For the $2qN_x$ solutions of the characteristic equation
\begin{equation}
    \det\mathcal{M}(\omega,\beta_x,e^{ik_y})=0~(k_y\in\mathbb{R}),\label{aaa}
\end{equation}
the condition for the GBZs is given by
\begin{equation}
    |\beta_{x,qN_x}|=|\beta_{x,qN_x+1}|,\label{GBZ_2d}
\end{equation}
where the solutions satisfy
\begin{equation}
    |\beta_{x,1}|\leq|\beta_{x,2}|\leq\cdots\leq|\beta_{x,2qN_x}|.
\end{equation}
We remark that the continuum bands obtained from the GBZ condition \eqref{GBZ_2d} reproduce the spectra under full OBCs in this case.

\subsection{Nonreciprocal-hopping Chern insulator}

In the following, we consider a nonlinear Chern insulator that exhibits the skin effect only in the $x$ direction. We demonstrate that the continuum bands obtained by combining Eqs.~\eqref{aaa} and \eqref{GBZ_2d} reproduce the spectra under full OBCs. We further show that a Chern number defined from the GBZs of auxiliary bands can predict the emergence of topological edge states.

\subsubsection{Model}

Suppose that the system is on an $L\times L$ square lattice. The Hamiltonian of the nonlinear Chern insulator is given by
\begin{eqnarray}
        &&H(\omega)=\sum_{n_x,n_y=1}^L\bm{c}_{(n_x,n_y)}^\dagger V(\omega)\bm{c}_{(n_x,n_y)}\nonumber\\
        &&+\sum_{n_x=1}^{L-1}\sum_{n_y=1}^L(\bm{c}_{(n_x,n_y)}^\dagger T_x\bm{c}_{(n_x+1,n_y)}+\bm{c}_{(n_x+1,n_y)}^\dagger T_x^\dagger\bm{c}_{(n_x,n_y)})\nonumber\\
        &&+\sum_{n_x=1}^L\sum_{n_y=1}^{L-1}(\bm{c}_{(n_x,n_y)}^\dagger T_y\bm{c}_{(n_x,n_y+1)}+\bm{c}_{(n_x,n_y+1)}^\dagger T_y^\dagger\bm{c}_{(n_x,n_y)}),\nonumber\\
        \label{2d_model}
\end{eqnarray}
where
\begin{eqnarray}
        &&V(\omega)=\begin{pmatrix}
            \alpha+m+\delta\omega&i\gamma_x\\i\gamma_x&\alpha-m-\delta\omega
        \end{pmatrix},\nonumber\\
        &&T_x=\frac12\begin{pmatrix}
            -t_x&-iv_x\\-iv_x&t_x
        \end{pmatrix},\quad T_y=\frac12\begin{pmatrix}
            -t_y&-v_y\\v_y&t_y
        \end{pmatrix},\nonumber\\
\end{eqnarray}
and $\bm{c}_{(n_x,n_y)}=(c_{(n_x,n_y),1},c_{(n_x,n_y),2})^\mathrm{T}$. We set $f(\omega)=\omega$ in Eq.~\eqref{rs_nep}, and assume that all the parameters take real values. This system preserves mirror symmetry with respect to the plane perpendicular to the $y$ direction, and the parameter $\gamma_x$ induces the skin effect in the $x$ direction. Meanwhile, the parameter $\delta$ represents the strength of the nonlinearity. We note that, when $\delta=0$, the system is described by a conventional eigenvalue equation with a non-Hermitian Hamiltonian.

\begin{figure}[t]
    \centering
    \resizebox{\linewidth}{!}{\input{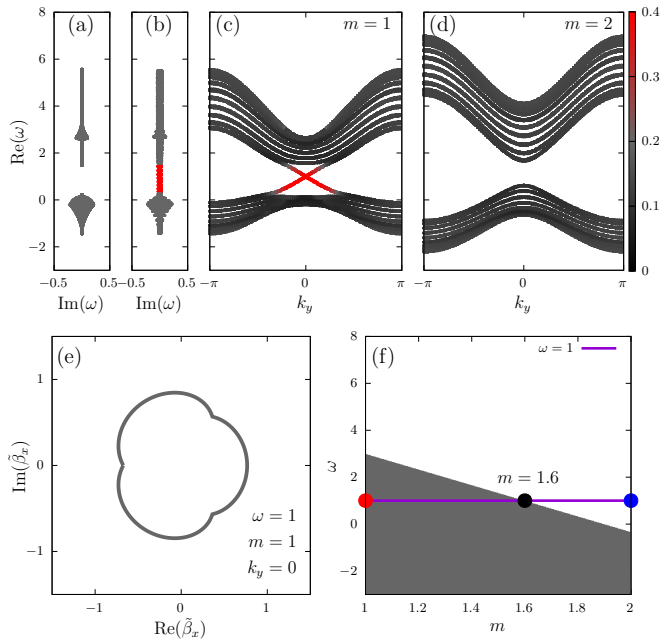}}
    \caption{Eigenvalues and Chern number of the nonlinear Chern insulator with nonreciprocal hopping amplitudes. (a) Continuum bands with $m=1$. (b) Spectra with $m=1$ under a full open boundary condition. The gray and red points represent the spectra of the bulk eigenstates and the edge states, respectively. (c) and (d) Spectra under the cylindrical geometry (open boundary conditions in the $x$ direction). The color bar means the biorthogonal inverse participation ratio defined in Eq.~\eqref{bi-IPR}. (e) Generalized Brillouin zones of the auxiliary bands $n=+$ and $n=-$. (f) Chern number \eqref{ChNum} for the lower band $n=-$. The Chern number is $-1$ and $0$ in the gray and white regions, respectively. The parameters are set to be $\delta = 0.3,\alpha = 1,\gamma_x = 0.2,v_x = 1.2,t_x = 0.9,v_y = 0.8$, and $t_y = 1$. The system size is $L=10$ in (b), (c), and (d).}
    \label{NRChInsulator}
\end{figure}

In this system, Eq.~\eqref{2d_M_k} with $\beta_y$ replaced by $e^{ik_y}~(k_y\in\mathbb{R})$ is written as
\begin{eqnarray}
    \mathcal{M}(\omega,\beta_x,e^{ik_y})&=&V(\omega)+\beta_x^{-1}T_x+\beta_xT_x^\dagger\nonumber\\
    &&+e^{-ik_y}T_y+e^{ik_y}T_y^\dagger-\omega\bm{1}_2.\label{Ch_aaa}
\end{eqnarray}
We note that this matrix satisfies $\mathcal{M}(\omega,e^{ik_x},e^{-ik_y})^\mathrm{T}=\mathcal{M}(\omega,e^{ik_x},e^{ik_y})$. Since the characteristic equation \eqref{aaa} is a quartic equation for $\beta_x$, the condition for the GBZs is given by
\begin{equation}
    |\beta_{x,2}|=|\beta_{x,3}|,\label{ChIns_GBZ}
\end{equation}
where the solutions of Eq.~\eqref{aaa} satisfy $|\beta_{x,1}|\leq|\beta_{x,2}|\leq|\beta_{x,3}|\leq|\beta_{x,4}|$. The combination of Eqs.~\eqref{Ch_aaa} and \eqref{ChIns_GBZ} allows us to calculate the continuum bands. By comparing Fig.~\ref{NRChInsulator}(a) and (b), we confirm that the continuum bands reproduce the bulk spectra under full OBCs. Thus, the non-Bloch framework discussed in Sec.~\ref{2d_NBF} is valid for describing a special class of 2D nonlinear systems.

\subsubsection{Bulk-boundary correspondence}

We confirm that the in-gap states emerge under a full OBC [Fig.~\ref{NRChInsulator}(b)] and the cylindrical geometry (OBC in the $x$ direction) [Fig.~\ref{NRChInsulator}(c)]. In contrast, the in-gap states disappear as the parameter $m$ increases [Fig.~\ref{NRChInsulator}(d)]. To investigate the localization behavior of the in-gap states, we consider the biorthogonal inverse participation ratio (bi-IPR)~\cite{Xiao2022}, which can distinguish the edge modes from the skin modes. Let $\mathcal{H}_{\rm cyl}(\omega,k_y)$ denote the system Hamiltonian under the cylindrical geometry. For the matrix $\mathcal{M}_{\rm cyl}(\omega,k_y)=\mathcal{H}_{\rm cyl}(\omega,k_y)-f(\omega)\bm{1}_{2L}$, we define $\ket{\chi^\mathrm{R}(\omega,k_y)}$ and $\bra{\chi^\mathrm{L}(\omega,k_y)}$ as
\begin{eqnarray}
    \mathcal{M}_{\rm cyl}(\omega,k_y)\ket{\chi^\mathrm{R}(\omega,k_y)}=\bm{0}, \\
    \bra{\chi^\mathrm{L}(\omega,k_y)}\mathcal{M}_{\rm cyl}(\omega,k_y)=\bm{0},
\end{eqnarray}
respectively. The bi-IPR is then defined as
\begin{equation}
    I_\text{bi}:=\left.\sum_{n_x=1}^L\sum_{\mu=1}^2\rho_{n_x,\mu}^2\right/\left(\sum_{n_x=1}^L\sum_{\mu=1}^2\rho_{n_x,\mu}\right)^2,\label{bi-IPR}
\end{equation}
where $\rho_{n_x,\mu}=|\chi^\mathrm{L}_{n_x,\mu}\chi^\mathrm{R}_{n_x,\mu}|$ is given by the components of $\ket{\chi^\mathrm{R}(\omega,k_y)}$ and $\bra{\chi^\mathrm{L}(\omega,k_y)}$. We find that the bi-IPRs of the in-gap states are larger than those of the bulk states [Figs.~\ref{NRChInsulator}(c) and (d)]. Physically, this means that the in-gap states can be interpreted as the edge states. We remark that the emergence of the edge states originates from the topological feature, as discussed below.

We consider an auxiliary system described by a conventional eigenvalue equation,
\begin{equation}
M(\omega)\ket{\Psi(\omega)}=\lambda(\omega)\ket{\Psi(\omega)},\label{zzz}
\end{equation}
where $M(\omega)$ is given in Eq.~\eqref{M_rs}, and $\lambda(\omega)$ is the auxiliary eigenvalue. We note that, when $\lambda(\omega)=0$, Eq.~\eqref{zzz} reduces to the nonlinear eigenvalue equation. The conventional non-Bloch band theory~\cite{Yao2018,Yokomizo2019,Yokomizo2020} allows us to calculate the auxiliary bands that reproduce the auxiliary spectra under full OBCs. We here introduce the characteristic equation as
\begin{equation}
\det[\lambda(\omega)\bm{1}_2-\mathcal{M}(\omega,\tilde{\beta}_x,e^{ik_y})]=0,\label{aux_ch_eq}
\end{equation}
where $\mathcal{M}(\omega,\tilde{\beta}_x,e^{ik_y})$ is given in Eq.~\eqref{2d_M_k}. Since this equation is a quartic equation for $\tilde{\beta}_x$, we can obtain the condition for the GBZs as
\begin{equation}
|\tilde{\beta}_{x,2}|=|\tilde{\beta}_{x,3}|.\label{yyy}
\end{equation}
As an example, we show the GBZs of both auxiliary bands for $k_y=0$ in Fig.~\ref{NRChInsulator}(e). We note that these GBZs completely overlap each other. By analogy with Ref.~\cite{Isobe2024}, a topological invariant defined in terms of the GBZ of auxiliary band can detect the presence or absence of the edge states in the original nonlinear system.

Let $n$ denote the band index in the following. Taking into account that the gap opens around $\lambda(\omega)=\alpha-\omega$, we label the auxiliary bands with $\rm{Re}(\lambda(\omega))<\rm{Re}(\alpha-\omega)$ and $\rm{Re}(\lambda(\omega))>\rm{Re}(\alpha-\omega)$ by $n=-$ and $n=+$, respectively. For $\bm{k}=(-i\ln\tilde{\beta}_x,k_y)$, we further denote the right and left eigenvectors of the matrix $\mathcal{M}(\omega,\bm{k})$ as $|\Phi_n^{\rm R}(\omega,\bm{k})\rangle$ and $\langle\Phi_n^{\rm L}(\omega,\bm{k})|$, respectively. We note that these vectors satisfy the biorthogonal condition $\bk{\Phi^\mathrm{L}_m(\omega,\bm{k})}{\Phi^\mathrm{R}_n(\omega,\bm{k})}=\delta_{mn}$ and the normalization convention $\langle\Phi_n^{\rm L}(\omega,\bm{k})|\Phi_n^{\rm L}(\omega,\bm{k})\rangle=1$. Following a previous work on a non-Hermitian Chern insulator~\cite{Yokomizo2023}, we define the Chern number of the auxiliary system as
\begin{equation}
    N_\text{Ch}^n(\omega):=\frac1{2\pi i}\oint_{T_n}\mathrm{d}\bm{k}\:\mathcal{B}_n(\omega,\bm{k}),\label{ChNum}
\end{equation}
where $\mathcal{B}_n(\omega,\bm{k})$ is the Berry curvature given by
\begin{eqnarray}
    \mathcal{B}_n(\omega,\bm{k})=\bk{\frac{\partial\Phi^\mathrm{L}_n(\omega,\bm{k})}{\partial \tilde{k}_x}}{\frac{\partial\Phi^\mathrm{R}_n(\omega,\bm{k})}{\partial k_y}}\nonumber\\
    -\bk{\frac{\partial\Phi^\mathrm{L}_n(\omega,\bm{k})}{\partial k_y}}{\frac{\partial \Phi^\mathrm{R}_n(\omega,\bm{k})}{\partial \tilde{k}_x}},
\end{eqnarray}
and $T_n$ represents the GBZ corresponding to the auxiliary band $n$.

By using the Fukui-Hatsugai-Suzuki method \cite{Fukui2005}, we numerically calculate the Chern number $N_\text{Ch}^-$ for the lower band $n=-$ [Fig.~\ref{NRChInsulator}(f)]. Since the gap in the continuum bands closes around $\omega=\alpha$, we can find the value of the parameter $m$ at which the Chern number switches its value. From Fig.~\ref{NRChInsulator}(f), for $\alpha=1$, the Chern number is $-1$ when $m<1.6$, which means that a nontrivial topological phase emerges in this parameter region. This result is consistent with the presence of the topological edge states [Figs.~\ref{NRChInsulator}(b) and (c)]. In contrast, when $m>1.6$, the Chern number becomes zero, and the edge states disappear [Fig.~\ref{NRChInsulator}(d)]. These results indicate that the topological phase transition occurs at $m=1.6$, which corresponds to the point at which the gap of the auxiliary bands closes around $\lambda(\omega)=0$. Therefore, these results establish the bulk-boundary correspondence between the nonzero Chern number and the emergence of the topological edge states.

\section{Summary and Outlook\label{summary_outlook}}

In summary, we develop the theoretical framework for calculating the continuum bands of tight-binding systems described by the nonlinear eigenvalue equation. We show how to calculate the continuum bands from the GBZs and demonstrate that they reproduce the spectra under OBCs. The GBZs allow us to analyze the localization behavior associated with the skin effect and various phenomena, such as the nonlinearity-induced skin effect. In the nonlinear Chern insulator with nonreciprocal hopping amplitudes, we establish the bulk-boundary correspondence between the Chern number defined from the GBZs of the auxiliary bands and the existence of the topological edge states.

The nonlinear eigenvalue equation considered here describes lattice systems, such as mechanical oscillators whose effective mass and effective stiffness depend on the eigenfrequencies~\cite{Fang2006,Yao2008,Huang2009,Huang2011,Lee2016,Srikanth2026} and electrical circuits that exhibit frequency-dependent gain~\cite{Zhang2026}. Accordingly, the non-Bloch band theory can provide a powerful tool for investigating topological properties of such systems in the presence of the skin effect. Furthermore, it would be possible to study the continuum bands of general 2D nonlinear systems in terms of the amoeba formulation \cite{Wang2024}. In addition, we expect that the non-Bloch band theory can be extended to a framework for analyzing continuous systems, which is analogous to previous studies on non-Hermitian systems~\cite{Yokomizo2022,Yokomizo2024,Hu2024}. For example, photonic crystals composed of metamaterials with frequency-dependent permittivity and permeability~\cite{Kittel1996,Jackson1999,Kuzmiak1994,Smith2002,Toader2004,Raghu2008,Mai2022} can be described within this framework. It should be intriguing to explore complex-frequency excitations~\cite{Kim2025,Long2026} in these photonic crystals by invoking the non-Bloch band theory.

\begin{acknowledgments}
We are grateful to Hikaru Matsumoto, Ryo Okugawa, Taiki Yoda, and Tsuneya Yoshida for valuable discussions. K.Y. was supported by JSPS KAKENHI Grant Numbers JP23K13027 and JP26K17049.
\end{acknowledgments}

\section*{Data Availability}

The data that support the findings of this article are not publicly available. The data are available from the authors upon reasonable request.

\appendix

\section{Condition for the generalized Brillouin zone\label{AppA}}

In this section, we show that, in the 1D nonlinear tight-binding system with Hamiltonian \eqref{fff}, the condition for the GBZs is given by Eq.~\eqref{GBZ}. The characteristic equation \eqref{M_ch_eq} is an algebraic equation for $\beta$ of $2qN$th degree, where $N$ and $q$ are the hopping range and the internal degrees of freedom, respectively. For a given $\omega$, the components of the eigenvector can be written in terms of the $2qN$ solutions of Eq.~\eqref{M_ch_eq} as
\begin{equation}
    \psi_{n,\mu}(\omega)=\sum_{j=1}^{2qN}\beta_j^{-n}\phi_\mu^{(j)}(\omega)~(n=1,\ldots,L,\;\mu=1,\ldots,q),\label{expansion}
\end{equation}
where $|\beta_1|\leq|\beta_2|\leq\dots\leq|\beta_{2qN-1}|\leq|\beta_{2qN}|$.

For the sake of simplicity, we impose the open boundary condition $\psi_{0,\mu}=\psi_{L+1,\mu}=0~(\mu=1,2,\dots,q)$. From Eq.~\eqref{expansion}, we obtain the simultaneous equations involving the $2q^2N$ coefficients $\phi_\mu^{(j)}(\omega)$. Meanwhile, the eigenvalue equation \eqref{k_ch_eq} determines the relations among the coefficients $\phi_1^{(j)}(\omega),\phi_2^{(j)}(\omega),\dots,\phi_\mu^{(j)}(\omega)$. Hence, the $2q^2N$ coefficients can be reduced to $2qN$ unknown variables. Suppose that $\phi_1^{(1)}(\omega),\phi_1^{(2)}(\omega),\dots,\phi_1^{(2qN)}(\omega)$ are taken as independent variables. The simultaneous equations can be then rewritten as $qN$ equations at the left end of the open chain,
\begin{equation}
    \sum_{j=1}^{2qN}F_l(\beta_j,\omega,\mathcal{S})\phi_1^{(j)}(\omega)=0,~(l=1,2,\ldots,qN),
\end{equation}
and $qN$ equations at the right end of the open chain,
\begin{equation}
    \sum_{j=1}^{2qN}\tilde{F}_l(\beta_j,\omega,\mathcal{S})\beta_j^{-L}\phi_1^{(j)}(\omega)=0,~(l=1,2,\ldots,qN),
\end{equation}
where $F_l$ and $\tilde{F}_l$ are functions of $\beta_j$, $\omega$, and the set of system parameters $\mathcal{S}$. We note that these functions do not depend on the system size $L$. By combining these equations, we obtain the condition for the existence of nontrivial solutions for $\phi_1^{(j)}$ as
\begin{equation}
    \begin{vmatrix}
        F_1(\beta_1,\omega,\mathcal{S})&\cdots&F_1(\beta_{2qN},\omega,\mathcal{S})\\
        \vdots&\ddots&\vdots\\
        F_{qN}(\beta_1,\omega,\mathcal{S})&\cdots&F_{qN}(\beta_{2qN},\omega,\mathcal{S})\\
        \tilde{F}_1(\beta_1,\omega,\mathcal{S})\beta_1^{-L}&\cdots&\tilde{F}_1(\beta_{2qN},\omega,\mathcal{S})\beta_{2qN}^{-L}\\
        \vdots&\ddots&\vdots\\
        \tilde{F}_{qN}(\beta_1,\omega,\mathcal{S})\beta_1^{-L}&\cdots&\tilde{F}_{qN}(\beta_{2qN},\omega,\mathcal{S})\beta_{2qN}^{-L}
    \end{vmatrix}=0.\label{obc_cond}
\end{equation}

The spectra of the open-boundary system can be obtained by solving Eq.~\eqref{obc_cond}. However, it is difficult to solve Eq.~\eqref{obc_cond} for sufficiently large system sizes. Nevertheless, we can investigate the asymptotic behavior of its solutions in the limit of a large system size. Equation~\eqref{obc_cond} can be rewritten as
\begin{equation}
    \sum_{P,Q}G(\beta_{i\in P},\beta_{j\in Q},\omega,\mathcal{S})\prod_{k\in Q}\beta_k^L=0,\label{obc_exp}
\end{equation}
where $P$ and $Q$ are two disjoint subsets of $\{1,2,\ldots,2qN\}$, and both have $qN$ elements. First, when $|\beta_{qN}|\neq|\beta_{qN+1}|$, there exists a single leading term on the left-hand side of Eq.~\eqref{obc_exp}, which is proportional to $(\beta_{qN+1}\beta_{qN+2}\cdots\beta_{2qN})^{L}$. Therefore, in the limit $L\rightarrow\infty$, Eq.~\eqref{obc_exp} reduces to
\begin{equation}
    G(\beta_{i\in P'},\beta_{j\in Q'},\omega,\mathcal{S})=0,
\end{equation}
with $P'=\{1,2,\ldots, qN\}$ and $Q'=\{qN+1,,qN+2,\ldots,2qN\}$. This condition leads to discrete sets of $\omega$. In contrast, when $|\beta_{qN}|=|\beta_{qN+1}|$, there are two leading terms on the left-hand side of Eq.~\eqref{obc_exp}: one proportional to $(\beta_{qN}\beta_{qN+2}\cdots\beta_{2qN})^{L}$ and the other proportional to $(\beta_{qN+1}\beta_{qN+2}\cdots\beta_{2qN})^{L}$. Hence, Eq.~\eqref{obc_exp} can be written as
\begin{equation}
    \left(\frac{\beta_{qN}}{\beta_{qN+1}}\right)^{L}=-\frac{G(\beta_{i\in P_0},\beta_{j\in Q_0},\omega,\mathcal{S})}{G(\beta_{i\in P_1},\beta_{j\in Q_1},\omega,\mathcal{S})},
\end{equation}
where $P_0=\{1,2,\ldots,qN\}, Q_0=\{qN+1,qN+2,\ldots,2qN\}, P_1=\{1,2,\ldots,qN-1,qN+1\}$, and $Q_1=\{qN,qN+2,\ldots,2qN\}$. This condition indicates that the relative phase between $\beta_{qN}$ and $\beta_{qN+1}$ can be changed almost continuously, which leads to continuum sets of $\omega$. Therefore, the eigenvalues under the OBC lie in the continuum sets of $\omega$ determined by $\beta_{qN}$ and $\beta_{qN+1}$. The continuum sets of $\beta_{qN}$ and $\beta_{qN+1}$ are formed by the condition $|\beta_{qN}|=|\beta_{qN+1}|$, which is nothing but  the condition for the GBZs.

\section{Topological aspect of the skin effect in the nonlinear system\label{AppB}}

In this section, we show that, when the winding number \eqref{windnum} takes nonzero values, the localized eigenstates emerge at one edge of the nonlinear system. We also show that, under the nonlinear pseudo-Hermiticity, the bulk eigenstates are localized at both edges of the system.

\subsection{Topological origin of the skin effect}
\label{AppB1}

We first consider a tight-binding system described by the nonlinear eigenvalue equation \eqref{rs_nep}. Solving the characteristic equation
\begin{equation}
    \det\mathcal{M}(\omega,e^{ik})=0
\end{equation}
enables us to calculate the Bloch bands, which reproduces the spectra under PBCs. We note that this equation is obtained by replacing $\beta$ with $e^{ik}~(k\in\mathbb{R})$ in Eq.~\eqref{M_ch_eq}. We recall that the winding number is defined as
\begin{equation}
    W(\omega_\text{ref}):=\frac1{2\pi i}\int_0^{2\pi}\mathrm{d}k\frac{\mathrm{d}}{\mathrm{d} k}\ln\det\mathcal{M}(\omega_\text{ref},e^{ik}),\label{def_windnum}
\end{equation}
where $\omega_\text{ref}\in\mathbb{C}$ is a reference point. Below, we show the bulk-edge correspondence between the nonzero winding number and the emergence of the skin modes. To this end, we define a Bloch Hamiltonian as
\begin{equation}
    \tilde{\mathcal{H}}(k):=\begin{pmatrix}
        O&\mathcal{M}(\omega_\text{ref},e^{ik})\\\mathcal{M}(\omega_\text{ref},e^{ik})^\dagger&O
    \end{pmatrix}.\label{ggg}
\end{equation}
This Bloch Hamiltonian preserves the chiral symmetry given by
\begin{equation}
    \varGamma\tilde{\mathcal{H}}(k)\varGamma^{-1}=-\tilde{\mathcal{H}}(k)\label{chiralsymaaa},
\end{equation}
where
\begin{equation}
    \varGamma=\sigma_z\otimes\bm{1}_q,~\sigma_z=\begin{pmatrix}
        1&0\\0&-1
    \end{pmatrix}.\label{hhh}
\end{equation}
Since the Hermitian Hamiltonian \eqref{ggg} belongs to class AIII in the Altland-Zirnbauer symmetry class \cite{Altland1997}, its topological phase is characterized by the winding number \eqref{def_windnum}~\cite{Ryu2010}.

As a consequence of the conventional bulk-boundary correspondence, when $W(\omega_m)\neq0$, zero-energy edge states emerge in the open-boundary Hamiltonian given by
\begin{equation}
    \tilde{H}:=\begin{pmatrix}
        O&M(\omega_m)\\M(\omega_m)^\dagger&O
    \end{pmatrix},\label{HermH_rs}
\end{equation}
where $M(\omega)$ is given by Eq.~\eqref{M_rs}, and $\omega_{m}$ are obtained by solving Eq.~\eqref{rs_eq}. We define the vectors $\ket{\psi^{\mathrm{R}}(\omega_m)}$ and $\ket{\psi^{\mathrm{L}}(\omega_m)}$ as
\begin{equation}
    M(\omega_m)\ket{\psi^\mathrm{R}(\omega_m)}=0,~M(\omega_m)^\dagger\ket{\psi^\mathrm{L}(\omega_m)}=0.
\end{equation}
The topological edge states of this Hamiltonian can then be constructed by
\begin{equation}
    \ket{\tilde{\psi}_+}=\begin{pmatrix}
        \ket{\psi^\mathrm{L}(\omega_m)}\\\bm{0}
    \end{pmatrix},~\ket{\tilde{\psi}_-}=\begin{pmatrix}
        \bm{0}\\\ket{\psi^\mathrm{R}(\omega_m)}
    \end{pmatrix}.
\end{equation}
The subscript $\pm$ denotes the chirality eigenvalue of $\Gamma=\sigma_z\otimes\bm{1}_{qL}$, namely, $\Gamma\ket{\tilde{\psi}_\pm}=\pm\ket{\tilde{\psi}_\pm}$. Since $\omega_m$ are the bulk spectra of the original nonlinear system, $\ket{\psi^\mathrm{R}(\omega_m)}$ represent the right eigenstates that are localized at one edge of the system. In other words, the nonzero winding number coincides with the emergence of the numerous localized states in the nonlinear system. We note that the sign of Eq.~\eqref{def_windnum} determines the localization direction of the skin modes.

\subsection{Localization property under the nonlinear pseudo-Hermiticity}
\label{AppB2}

We next investigate the localization property of the skin mode under the nonlinear pseudo-Hermiticity. As shown below, the bulk eigenstates are localized at both edges of the system. In this case, we recall that the matrix $\mathcal{M}(\omega,e^{ik})$ satisfies
\begin{equation}
    \mathcal{M}(\omega^\ast,e^{ik})=\mathcal{U}\mathcal{M}(\omega,e^{ik})^\dagger\mathcal{U},
\end{equation}
where $\mathcal{U}$ is a $q\times q$ unitary and Hermitian matrix. For a given reference point $\omega_{\rm ref}\in\mathbb{C}$, we define a Bloch Hamiltonian as
\begin{equation}
    \tilde{\mathcal{H}}_\mathcal{U}(k):=\begin{pmatrix}
        O&\mathcal{U}\mathcal{M}(\omega_\text{ref},e^{ik})\\\mathcal{M}(\omega_\text{ref},e^{ik})^\dagger\mathcal{U}&O
    \end{pmatrix}.
\end{equation}
Since this Bloch Hamiltonian preserves the chiral symmetry
\begin{equation}
    \varGamma\tilde{\mathcal{H}}_\mathcal{U}(k)\varGamma^{-1}=-\tilde{\mathcal{H}}_\mathcal{U}(k)\label{chiralsymaaa},
\end{equation}
where $\varGamma$ is given by Eq.~\eqref{hhh}, it is characterized by a $\mathbb{Z}$ topological invariant,
\begin{equation}
    W_\mathcal{U}(\omega_\text{ref}):=\frac1{2\pi i}\int_0^{2\pi}\mathrm{d}k\frac{\mathrm{d}}{\mathrm{d}k}\ln\det[\mathcal{U}\mathcal{M}(\omega_\text{ref},e^{ik})].
\end{equation}
We note that this invariant is equivalent to the winding number \eqref{def_windnum} because $\det[\mathcal{U}\mathcal{M}(\omega_\text{ref},e^{ik})]=\det\mathcal{M}(\omega_\text{ref},e^{ik})$.

Under the nonlinear pseudo-Hermiticity, the matrix $M(\omega)$ given in Eq.~\eqref{M_rs} satisfies
\begin{equation}
    M(\omega^\ast)=UM(\omega)^\dagger U,\label{iii}
\end{equation}
where $U$ is a $qL\times qL$ unitary and Hermitian matrix. From the conventional bulk-edge correspondence, when $W_\mathcal{U}(\omega_m)\neq0$, there exist zero-energy edge states of the Hamiltonian given by
\begin{equation}
    \tilde{H}_U:=\begin{pmatrix}
        O&UM(\omega_m)\\M(\omega_m)^\dagger U&O
    \end{pmatrix},\label{jjj}
\end{equation}
where $\omega_m$ is obtained by solving Eq.~\eqref{rs_eq}. By using Eq.~\eqref{iii}, we can obtain the topological edge states of the system as
\begin{equation}
    \ket{\tilde{\psi}_{U,+}}=\begin{pmatrix}
        \ket{\psi(\omega_m^\ast)}\\\bm{0}
    \end{pmatrix},~\ket{\tilde{\psi}_{U,-}}=\begin{pmatrix}
        \bm{0}\\\ket{\psi(\omega_m)}
    \end{pmatrix},\label{mmm}
\end{equation}
where $\ket{\psi(\omega_m)}$ satisfies
\begin{equation}
    M(\omega_m)\ket{\psi(\omega_m)}=0,
\end{equation}
and the meaning of the subscript $\pm$ is the same as in Sec.~\ref{AppB1}. We here remark that the topological edge states with opposite chirality are localized at the left and right edges of the system, respectively. This means that the bulk eigenstates $|\psi(\omega_m)\rangle$ and $|\psi(\omega_m^\ast)\rangle$ of the nonlinear system are localized at the edges opposite to each other. Therefore, systems with the nonlinear pseudo-Hermiticity exhibit the bidirectional localization behavior.

\bibliography{ref.bib}

\end{document}